\documentclass[prl,twocolumn,showpacs,groupaddress]{revtex4}
\usepackage{graphicx,amsfonts,amsmath,amssymb}
\usepackage{epsfig}
\usepackage{color}
\usepackage{epstopdf}
\usepackage{txfonts}

\begin{document}
\title{Quantum and thermal effects in dark soliton formation and dynamics in a 1D Bose gas}
\author{A. D. Martin}
\affiliation{School of Mathematics, University of Southampton, Southampton SO17 1BJ, United Kingdom}
\author{J. Ruostekoski}
\affiliation{School of Mathematics, University of Southampton, Southampton SO17 1BJ, United Kingdom}

\date{\today}
\begin{abstract}
We numerically study the imprinting and dynamics of dark solitons in a bosonic atomic
gas in a tightly-confined one-dimensional harmonic trap both with and without an optical lattice.
Quantum and thermal fluctuations are synthesized within the truncated Wigner approximation in the quasi-condensate description. We track the soliton coordinates and calculate position and velocity uncertainties. We find that the phase fluctuations {\em lower} the classically predicted soliton speed and seed instabilities. Individual runs show interactions of solitons with sound waves, splitting and disappearing solitons.
\end{abstract}

 \maketitle

Dark solitons, or phase kinks, are commonly studied excitations in atomic Bose-Einstein condensates (BECs) and in nonlinear optics \cite{KIV98}, and are reminiscent of robust nonlinear solitary waves emerging in numerous other physical systems. The main research emphasis has been on classical mean-field properties of weakly interacting systems that can be accurate modeled by the nonlinear Schr\"odinger or the Gross-Pitaevskii equation (GPE) \cite{anglin}. In BEC experiments dark solitons were prepared by imprinting a sharp phase jump on the atomic cloud by optical potentials \cite{Burger_PRL_1999}, by ultraslow light \cite{DUT01}, or by merging two BECs \cite{Weller}, and the two-dimensional (2D) and 3D soliton dynamics were found to be unstable to perturbations perpendicular to the soliton velocity (snake instability). Classically solitons can also decay due to sound emission generated by an anharmonic trapping potential \cite{Burger_PRA_2002,Kevrekidis_PRA_2003,Parker_JPhysB_2004,Theocharis_MCSim_2005}.

\begin{figure}[tbp]
\includegraphics[width=\columnwidth]{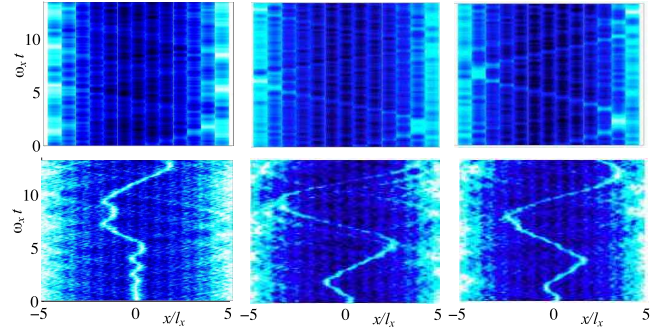}\vspace{-3mm}
\caption{(Color online) Atom densities of individual stochastic realizations of soliton dynamics in a combined harmonic trap and optical lattice at $T=0$. Top: Different single-shot outcomes for the same system obtained from the Wigner function $\psi_W$ projected to the lowest energy band in a lattice. The imprinted phase in the corresponding classical case $\phi_c=1.3$ and $N_{\mbox{\scriptsize tot}}=240$. The solitons disappear or split. Bottom: The Wigner density for $\phi_c=\pi$,  $N_{\mbox{\scriptsize tot}}=900$ (left), and for two different stochastic trajectories (both with $\phi_c=\pi$, $N_{\mbox{\scriptsize tot}}=440$) that start to oscillate in opposite directions (center and right).} \label{Lowest_Plots}
\end{figure}

In atomic BECs solitons are manifestations of quantum mechanics on a macroscopic (or mesoscopic) scale. Soliton properties are sensitive to fluctuations of the BEC wavefunction and provide an ideal system to study the emergence of classical physics from interacting quantum many-body physics \cite{Anglin_NaturePhys_2008}.  In recent experiments bosonic atoms were confined in tight elongated 1D traps that, together with optical lattices, significantly enhance quantum fluctuations due to stronger effects of interactions \cite{Fertig_PRL_2005,KIN06}, rendering classical descriptions invalid. In this Letter we study quantum properties of solitons by considering the phase imprinting process and the resulting nonequilibrium quantum dynamics of phase kinks in such systems by including the enhanced quantum and thermal fluctuations of the atoms within the truncated Wigner approximation (TWA) \cite{DRU93,Isella_PRA_2006,Isella_PRL_2005}. In 1D the snake instability is suppressed and quantum features of solitons are easily distinguishable. Individual stochastic realizations of TWA represent possible outcomes of single experimental runs revealing jittering oscillatory motion, splitting and disappearing solitons (Fig.~\ref{Lowest_Plots}), while the ensemble averages calculated from TWA produce a quantum statistical description of the soliton dynamics over its entire trajectory. Numerical tracking of soliton coordinates at different times allows us to evaluate the position and velocity uncertainties of the phase kinks. We find them rapidly growing as a function of quantum and thermal depletion, but, surprisingly, observe the average soliton speed being reduced due to enhanced fluctuations as a result of the nonlinear dependence of the soliton speed upon its phase distribution. In a lattice quantum and thermal fluctuations seed the dynamical instabilities of the corresponding classical system--in extreme cases the soliton motion is solely generated by quantum fluctuations. Fluctuations also mitigate the unstable trapping of solitons in individual lattice sites by stimulating sound emission and decreasing the position uncertainty of the solitons.

TWA has proved particularly useful for the studies of dissipative quantum dynamics of bosonic atoms in 1D optical lattices when the atom filling factor in the lattice is not too small. For example, it qualitatively produced the experimentally observed damping rate \cite{Fertig_PRL_2005} of the center-of-mass motion of the atom cloud even for significant ground state depletion \cite{Isella_PRL_2005}. Here we show that the quantum statistics of the atoms in TWA can be synthesized according to the quasi-condensate description, analogous to the long-wavelength limit of the Luttinger liquid theory, providing an improved representation of TWA--phase kinks are particularly sensitive to the resulting enhanced phase fluctuations. By varying the effective interaction parameter we can study a smooth transition of soliton dynamics from a classical to a strongly fluctuating regime. An additional advantage of TWA over quantum field-theoretical soliton descriptions is that it can more easily incorporate excitations of the system far from the thermal equilibrium as typical in soliton imprinting experiments. In nonlinear optics TWA was used to analyze solitons in fibers \cite{optsoliton}. In previous theoretical dark soliton studies in BECs thermal atoms were shown to damp the soliton motion \cite{Jackson_PRA_2007} and quantum fluctuations \cite{KRU09} fill the soliton core \cite{Dziarmaga_JphysB_2005,Mishmash_unpublished_2008} or affect the statistics \cite{Negretti_PRA_2008_b}.

We assume the atom dynamics to be restricted to 1D by a strong radial confinement, so that TWA dynamics follow from an ensemble of stochastic fields $\psi_{W}$ satisfying \cite{DRU93}
\begin{equation}
i\hbar \frac{\partial}{\partial
t}\psi_{W}=\big(-\frac{\hbar^{2}}{2m}\frac{\partial^{2}}{\partial x^{2}}+V +g_{\mbox{\scriptsize1D}}N_{\mbox{\scriptsize tot}}
|\psi_{W}|^{2}\big)\psi_{W},
\label{GPE}
\end{equation}
with $g_{1D}=2\hbar \omega_{r}a$, where $a$ denotes the $s$-wave scattering length and $\omega_{r}$ the radial trap frequency. Equation (\ref{GPE}) formally coincides with GPE, but here $\psi_{W}(x,t)$ is a stochastic phase-space representation of the full field operator. The noise is included in the initial conditions of $\psi_{W}$ that synthesize the quantum statistical correlation functions of the atoms and generate the fluctuations of TWA dynamics. As phase kinks represent defects sensitive to phase fluctuations, we generate the initial state noise for the bosonic field operator $\hat{\psi}(x,0)$ carefully using a 1D quasi-condensate description \cite{Mora_PRA_2003} by introducing the density $\delta{\hat{\rho}}(x)$ and phase $\hat{\varphi}(x)$ operators
%
\begin{equation}
\begin{split}
\hat{\psi}(x,0) &=\sqrt{\rho_{0}(x)+\delta{\hat{\rho}}(x)}\exp(i\hat{\varphi}(x)),\label{Quasi}\\
\hat{\varphi}(x) &=\frac{-i}{2\sqrt{\rho_{0}(x)}}\sum_{j}\left(\varphi_{j}(x) \hat{\alpha}_{j}-\varphi_{j}^{*}(x)\hat{\alpha}_{j}^{\dagger}\right),\\
\delta\hat{\rho}(x) & = \sqrt{\rho_{0}(x)}\sum_{j}\left(\delta\rho_{j}(x)\hat{\alpha}_{j}+ \delta\rho_{j}^{*}(x)\hat{\alpha}_{j}^{\dagger}\right)\,,
\end{split}
\end{equation}
%
where $\varphi_{j}(x)=u_{j}(x)+v_{j}(x)$ and $\delta\rho_{j}(x)=u_{j}(x)-v_{j}(x)$ are given in terms of the solutions to the Bogoliubov equations, $u_{j}(x)$ and $v_{j}(x)$ \cite{Isella_PRA_2006}. Here $\rho_{0}=\bar{N}_{0}|\psi_{0}(x)|^{2}$ and $\psi_{0}(x)$ is the ground state wavefunction with $\bar{N}_{0}$ particles.

In TWA the initial conditions are given by Eq.~(\ref{Quasi}), with the operators $(\hat{\alpha}_{j}^{\dagger},\hat{\alpha}_{j})$ replaced by complex variables $(\alpha_{j}^{*},\alpha_{j})$ sampled from the relevant Wigner distribution--in this case representing ideal harmonic oscillators in a thermal bath \cite{Isella_PRA_2006}, such that $\langle \alpha_{j}^{*}\alpha_{j} \rangle_{W}=\bar{n}_{j}+\frac{1}{2}$. Here $\frac{1}{2}$ results from the Wigner representation that returns symmetrically ordered expectation values. We consider the total atom number $N_{\mbox{\scriptsize tot}}$ to be fixed at each stochastic run. The ground state population $N_0$ for a particular run may then be defined with reference to the depleted population $N_{\mbox{\scriptsize nc}}$ of the same run
\begin{equation}\label{depletion}
N_{\mbox{\scriptsize nc}}=\int dx \sum_j \left[ \left(|\alpha_{j}|^{2}-1/2\right)\left(|u_{j}(x)|^{2}+|v_{j}(x)|^{2}\right)+|v_{j}(x)|^{2}\right],
\end{equation}
so that for each run we set $N_{0}=N_{\mbox{\scriptsize tot}}-N_{\mbox{\scriptsize nc}}$. Here $N_{\mbox{\scriptsize nc}}$ fluctuates at each realization around its average value $\bar{N}_{\mbox{\scriptsize nc}}$, obtained from Eq.~(\ref{depletion}) with the substitution $|\alpha_{j}|^{2}\rightarrow \langle\hat\alpha_j^\dagger\hat\alpha_j\rangle+1/2$, and $N_0$ fluctuates around the average value $\bar{N}_0=N_{\mbox{\scriptsize tot}}-\bar{N}_{\mbox{\scriptsize nc}}$. This provides a simple description to sample the ground state mode operator. In the rest of the paper, when specifying the depletion we refer to the expectation value $\bar{N}_{\mbox{\scriptsize nc}}/N_{\mbox{\scriptsize tot}}$.
We consider the effects of varying $\bar{N}_{\mbox{\scriptsize nc}}/N_{\mbox{\scriptsize tot}}$, either by changing $T$ or $g_{\mbox{\scriptsize 1D}}/N_{\mbox{\scriptsize tot}}$ whilst keeping the nonlinearity constant at $g_{\mbox{\scriptsize 1D}}N_{\mbox{\scriptsize tot}}=100\hbar\omega_{x}l_{x}$, where $\omega_{x}$ denotes the axial trap frequency and $l_{x}=\sqrt{\hbar/m\omega_{x}}$. Here $g_{\mbox{\scriptsize 1D}}/N_{\mbox{\scriptsize tot}}\propto\gamma_{\rm int}$ where the effective 1D interaction parameter $\gamma_{\rm int}=m g_{1D}/\hbar^2 n$ and $n$ is the 1D atom density.
For comparison, in the recent experiment \cite{Fertig_PRL_2005}, atoms were confined to an array of 1D tubes with $g_{\mbox{\scriptsize 1D}}N_{\mbox{\scriptsize tot}}\simeq 320\hbar\omega_{x}l_{x}$ and $N_{\mbox{\scriptsize tot}}\simeq70$ in the central tube. The ratio $g_{\mbox{\scriptsize 1D}}/N_{\mbox{\scriptsize tot}}$ can be experimentally managed, e.g., by adjusting the radial trap frequency or the scattering length as $N_{\mbox{\scriptsize tot}}$ is varied.
\begin{figure}[tbp]
\includegraphics[width=\columnwidth]{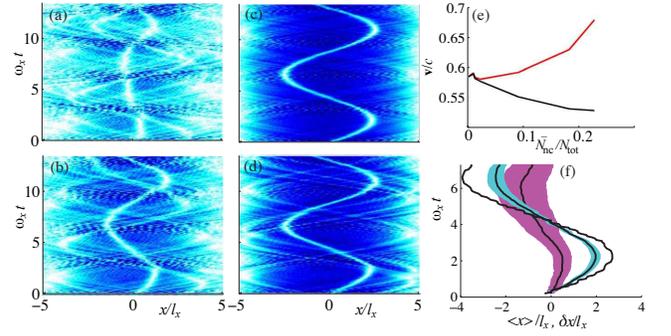}
\vspace{-8mm}\caption{(Color online) Soliton dynamics in a harmonic trap without a lattice. 
(a-d) The Wigner density $|\psi_{w}(x,t)|^{2}$ for individual stochastic realizations (that represent possible outcomes of single experimental realizations) with the same $g_{\mbox{\scriptsize 1D}}N_{\mbox{\scriptsize tot}}$ and $\phi_{c}=2$; (a-c) At $T=0$ for $N_{\mbox{\scriptsize tot}}=50,100,900$, corresponding to  $\bar{N}_{\mbox{\scriptsize nc}}/N_{\mbox{\scriptsize tot}}=0.2, 0.1, 0.01$, respectively. The emitted sound pulses from the imprinting process interact more with the soliton at low atom numbers and large quantum fluctuations; (d)  $N_{\mbox{\scriptsize tot}}=900$, $k_{B}T/\hbar\omega_{x}=22$ with   $\bar{N}_{\mbox{\scriptsize nc}}/N_{\mbox{\scriptsize tot}}=0.3$. Thermal effects are weaker even with large thermal depletion; (e) Soliton velocity at $T=0$ that would follow from $\cos(\langle\hat{\phi}\rangle/2)$ for the expectation value of the imprinted phase $\langle \hat\phi \rangle$ (red/top curve) and the expectation value of the velocity obtained from  $\langle \cos(\hat\phi/2) \rangle$ (black curve) immediately after the imprinting vs depletion; (f) The ensemble
averages of the soliton positions $\langle x\rangle$ (solid lines) and the quantum mechanical position uncertainties $\delta x$ (shaded regions) for $N_{\rm tot}=8000,440,50$ (curves with decreasing amplitudes) with the same nonlinearity $g_{\mbox{\scriptsize 1D}}N_{\mbox{\scriptsize tot}}$, $\phi_{c}=2$, and $T=0$. At $N_{\mbox{\scriptsize tot}}\simeq8000$, $\delta x$ is negligible. Quantum fluctuations increase $\delta x$ and soliton damping, and decrease the speed.} \label{Wigner_Plots}
\end{figure}

We numerically study the imprinting process of phase kinks in a 1D bosonic
gas and the resulting TWA quantum dynamics (\ref{GPE}).
We consider the experimental imprinting method \cite{Burger_PRL_1999}, where a soliton is generated by applying an additional constant `light-sheet potential', of value $V_{\phi}$ to half of the atom cloud, for time $\tau$, so that the external potential in Eq.~(\ref{GPE}) reads
\begin{equation}
V= m\omega_{x}^{2}x^{2}/2+sE_{r}\sin^2( \pi x/d)+\theta(\tau-t)\theta(x)V_{\phi}\,.
\end{equation}
Here $s$ denotes the lattice strength in the units of lattice photon recoil energy $E_{r}=\hbar^{2}\pi^{2}/2md^{2}$ with the lattice spacing $d$. In the corresponding classical case the light sheet imprints a phase jump of $\phi_c=V_{\phi}\tau/\hbar$ at $x=0$, preparing a dark soliton. The imprinted solitons were consequently free to evolve in a harmonic trap (Fig.~\ref{Wigner_Plots}) and in a combined harmonic trap and lattice (Fig.~\ref{Lowest_Plots}). The corresponding classical soliton (governed by GPE) oscillates in a harmonic trap ($s=0$) at the frequency $\omega_{x}/\sqrt{2}$ \cite{anglin} with
the initial velocity ${\bf v}/c=\cos(\phi_c/2)$, depending on the imprinted phase $\phi_c$, where $c(x)=\sqrt{g_{\mbox{\scriptsize 1D}}N_{\mbox{\scriptsize tot}}|\psi(x)|^2/m}$ is the speed of sound. The soliton is stationary (dark) for $\phi_c=\pi$, with a zero density at the kink. Other phase jumps produce moving (grey) solitons, with non-vanishing densities $n_{s}=n\cos^{2}(\phi_{c}/2)$ at the phase kink, where $n$ is the background density, and $|{\bf v}|\rightarrow c$ for
$\phi_c\rightarrow 0$. Hence, the soliton speed and depth can be controlled by $\tau$ or  $V_{\phi}$ \cite{Burger_PRL_1999}.

\begin{figure}[tbp]
\includegraphics[width=\columnwidth]{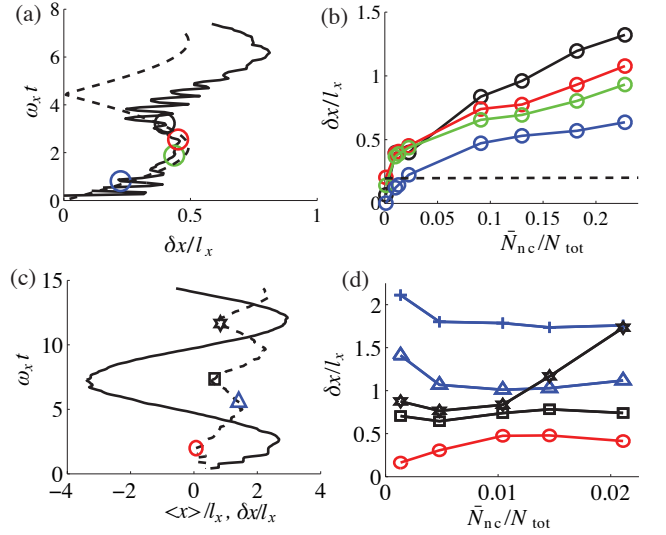}
\vspace{-9mm}\caption{(Color online) (a) The evolution of the soliton position uncertainty $\delta x$ in a harmonic trap as numerically tracked from the soliton coordinates at $T=0$ for $N_{\mbox{\scriptsize tot}}=440$ ($\bar{N}_{\mbox{\scriptsize nc}}/N_{\mbox{\scriptsize tot}}\simeq 0.02$; solid line) and the position uncertainty of an ensemble of sinusoids with the corresponding initial velocity distribution (dotted line), with $\delta x$ quickly becoming larger than predicted by the uncertainty in the initial velocity alone; (b) $\delta x$ at four times $t\leq5/\omega_{x}$ [as marked in (a)] vs depletion, obtained by varying $g_{\mbox{\scriptsize 1D}}/N_{\mbox{\scriptsize tot}}$ for constant $g_{\mbox{\scriptsize 1D}}N_{\mbox{\scriptsize tot}}$. The horizontal dotted line indicates the soliton core size at the trap center; (c) The quantum expectation value of the soliton position $\langle x \rangle$ (solid line) and its uncertainty $\delta x$ (dotted line) in a combined harmonic trap and optical lattice for $s=1$, $T=0$, $N_{\mbox{\scriptsize tot}}=8000$ ($\bar{N}_{\mbox{\scriptsize nc}}/N_{\mbox{\scriptsize tot}}\simeq 0.001$);
(d) $\delta x$ [at times indicated with corresponding symbols in (c)] at $T=0$. In all cases the soliton positions are calculated before solitons break or split in individual trajectories. } \label{Trajectory_Plots}
\end{figure}

In quantum case, soliton trajectories in TWA fluctuate between different realizations. Individual stochastic realizations of $|\psi_W|^2$ in a harmonic trap ($s=0$) for different atom numbers and temperatures are shown in Fig.~\ref{Wigner_Plots}(a-d) that represent possible experimental observations (quantum measurements) of single runs. The phase imprinting creates sound waves \cite{Burger_PRL_1999} that more strongly interact with the individual soliton trajectories at low atom numbers (for a fixed nonlinearity). We also show the relative effects of thermal versus quantum depletion. We find the effect of even a weak $T=0$ quantum depletion clearly more significant than thermal depletion (evaluated for large atom numbers for which $T=0$ depleted fraction is negligible), in both increased deviations from the classical sinusoidal oscillations and in damping.  Fluctuations `damp' the dipolar soliton motion such that the oscillation amplitude is increased, but the frequencies fluctuate only slightly around the classical value of  $\omega_{x}/\sqrt{2}$. In TWA we can ensemble average hundreds of stochastic realizations in order to obtain quantum statistical correlations of the soliton dynamics. We numerically track the position of the kink at different times in individual realizations and calculate the quantum mechanical expectation values for the soliton position $\langle x \rangle$ and its uncertainty $\delta x=\sqrt{\langle x^{2}\rangle-\langle x\rangle^2}$. The Wigner probabilities can be transformed to normally ordered expectation values by subtracting half-an-atom per phonon mode.
In Fig.~\ref{Wigner_Plots}(f) we show $\langle x \rangle$ and $\delta x$ for systems with different values of ground state depletion, clearly demonstrating increasing $\delta x$ due to quantum fluctuations [also shown in Fig.~\ref{Trajectory_Plots}(b)]. In
Fig.~\ref{Trajectory_Plots}(a) we find that $\delta x$ at later times is larger than predicted by the initial velocity uncertainty alone. We also fit the individual soliton trajectories with the curve $x(t)=f(t)\exp(-\gamma t)$, where $f(t)$ is an undamped sinusoid and find the quantum expectation value for the damping rate  $\bar{\gamma}\simeq -0.057\omega_{x}$ for $T=0$, $N_{\mbox{\scriptsize tot}}=900$, increasing with $T$ up to $\bar{\gamma} \simeq-0.073\omega_{x}$ at $T=22\hbar\omega/k_B$, corresponding to 30\% depleted fraction. At $T=0$, $\bar\gamma$ is reduced with increasing $N_{\mbox{\scriptsize tot}}$, with $\bar\gamma\simeq -0.023\omega_x$ at $N_{\mbox{\scriptsize tot}}\simeq 8000$. The qualitative features of the finite-temperature damping
for large atom numbers (when quantum fluctuations are negligible at $T=0$) compares to the system analyzed in a different regime in Ref.~\cite{Jackson_PRA_2007}, where noticeable damping only occurs over many oscillation periods.

We find that the fluctuations of the phase jump across the soliton dramatically affect the expectation value and the uncertainty of the soliton velocity [Fig.~\ref{Wigner_Plots}(e,f)]. Similarly to previous studies \cite{Dziarmaga_JphysB_2005}, we find that quantum depletion causes filling of the soliton core in the ensemble average of atom densities. The classical relation between the soliton depth and speed suggests the depletion might also increase the soliton speed, but we show that, in fact, the opposite is true. The quantum expectation value of the phase jump $\langle\hat\phi\rangle$ decreases due to quantum fluctuations. However, the classical expression for the soliton speed $|\cos(\phi/2)|$ has a negative curvature, so that for a symmetric phase distribution the quantum expectation value $|\langle\cos(\hat\phi/2)\rangle|$ is always smaller than the speed resulting from the expectation value of the phase jump $|\cos(\langle \hat\phi/2 \rangle)|$, and will decrease as the width of the phase distribution increases. For instance, if the phase distribution has a small width $\Delta\phi$, the speed is approximately reduced by the factor $1-\Delta\phi^2/32$, since the average of $\cos\left[(\phi\pm\Delta\phi/2)/2\right]$ approximately yields
$\cos(\phi/2)\left[1-\Delta\phi^{2}/32\right]$. Fig.~\ref{Wigner_Plots}(e) shows that despite  $\langle\hat\phi\rangle$ decreasing with increasing ground state depletion, the average initial speed decreases due to the broader phase distribution dominating over the effect of decreasing $\langle \hat\phi\rangle$. We find that the formula $|\langle\cos(\hat\phi/2)\rangle|$ gives good agreement with the average initial speed, but underestimates (overestimates) it by a few percent at large (small) atom numbers. 

Different approximations in the stochastic sampling of the initial quantum statistical correlation functions affect the soliton dynamics. In the Bogoliubov approximation \cite{Isella_PRA_2006} the imprinted phase distribution is slightly narrower than in the quasi-condensate description. For small depletion the difference in dynamics between the two approximations is negligible, but for quantum depletion $\bar{N}_{\mbox{\scriptsize nc}}/N_{\mbox{\scriptsize tot}}\simeq 0.2$, it is typically about 8\%, reaching 30\% at later times when $\delta x$ becomes large.

We next consider phase kink dynamics in a combined harmonic trap and an optical lattice. In classical soliton dynamics a lattice introduces damping and instabilities at low velocities \cite{Kevrekidis_PRA_2003,Parker_JPhysB_2004}. We consider a lattice with $\pi l_x/d=4$, corresponding to about 21 occupied sites. Shallow lattices ($s=0.25$) show no qualitative effects in the soliton dynamics, but this changes at about $s=0.5$. In Fig.~\ref{Lowest_Plots} individual stochastic realizations at $s=1$ display strong effects of quantum fluctuations exhibiting rich variation that is not always easily captured by quantum mechanical ensemble averaging: fast solitons may disappear, or split into multiple solitons at later times, even for small depletion $\bar{N}_{\mbox{\scriptsize nc}}/N_{\mbox{\scriptsize tot}}=0.04$. We also calculate $\delta x$ (Fig.~\ref{Trajectory_Plots}) that becomes very large before the break-up or disappearance.

Stationary solitons at the harmonic trap minimum in a lattice are at an unstable equilibrium \cite{Kevrekidis_PRA_2003}, but even at low initial velocities first remain trapped in the central site before starting to oscillate \cite{Theocharis_MCSim_2005} (Fig.~\ref{Lowest_Plots}), as it takes time for the soliton to lose energy via sound emission and move into the neighboring site. In quantum simulations we find the motion seeded by fluctuations even for $\phi_c=\pi$ when the expectation value of the initial velocity vanishes, corresponding to large fluctuations for the soliton tunneling time out of the central site. For $\phi_c=\pi$ and $N_{\mbox{\scriptsize tot}}=900$ at $T=0$ the tunneling time and its uncertainty are of the order of the trap period. At low initial speeds quantum fluctuations may even change the direction of the initial velocity, generating large variations between individual trajectories and completely dominating the dynamics [Fig.~\ref{Lowest_Plots} (bottom)]. Large soliton uncertainties were also seen in systems of small atom numbers in the tight-binding limit \cite{Mishmash_unpublished_2008}.

Phase kinks may also become trapped in the outer wells at the turning points of their trajectories [Fig.~\ref{Lowest_Plots}]. The increased trapping time in the outer lattice sites accompanies larger fluctuations in tunneling times, so that $\delta x$ increases here for systems with smaller depletion. Figure \ref{Trajectory_Plots}(d) shows that the uncertainties in position are initially greatest in the low atom number and high temperature cases, due to the broad initial velocity distribution, but the situation is reversed after the first turning point in the soliton trajectories. Hence the surprising result that phase kinks in systems with a small ground state depletion exhibit larger position uncertainties here than those in systems with a large depletion. The increased sound emission is also evidenced by the damping rate. A soliton with an imprinted phase jump of $\phi_c=2.6$ has a classical damping rate $-0.08\omega_{x}$, which increases to the average quantum value $\bar\gamma\simeq-0.1\omega_{x}$ as the number of atoms is reduced to $N_{\mbox{\scriptsize tot}}=900$ at $s=1$, $T=0$, corresponding to $\bar{N}_{\mbox{\scriptsize ns}}/N_{\mbox{\scriptsize tot}}=0.01$. Faster solitons only experience weak damping (e.g. $\phi_c=2.0$ with $\bar\gamma=-0.04\omega_{x}$) with little effect due to similar depletion.

In conclusion, we showed that solitons can provide an ideal dynamical observable to probe underlying quantum fluctuations. The TWA phase-space description can incorporate a very large number of degrees of freedom; the dissipative dynamics of solitons or the entire atom cloud \cite{Isella_PRL_2005} emerge from a microscopic treatment of the unitary quantum evolution without any explicit damping terms. The advantage is that the frequently problematic separation of quantum dynamics to `system' degrees of freedom and `environment' \cite{Anglin_NaturePhys_2008} is not needed and classical physical observables naturally and unambiguously emerge in the formalism. Here our numerical tracking of the soliton coordinates provides outcomes of single-shot measurements of soliton trajectories as well as a precise quantum statistical description of the soliton position and velocity.

We acknowledge discussions with Y.\ Castin and financial support from EPSRC.

\end{document}